\documentclass[a4paper]{jpconf}
\usepackage{geometry,amsmath,amsfonts}
\usepackage{slashed}
\usepackage{epsfig}
\usepackage{latexsym}
\usepackage{graphicx}
\usepackage{epstopdf}  
\usepackage[justification=RaggedRight]{caption}
\usepackage{amssymb}
\usepackage{subfig}
\usepackage{color}
\usepackage{multirow}
\usepackage{color}
\usepackage{rotating}
\usepackage{ifthen}
\usepackage{epsfig}
\usepackage{graphics}
\usepackage{xcolor}
\usepackage{enumitem}
\begin{document}
\title{Dark Matter and Baryon Asymmetry production from out-of-equilibrium decays of Supersymmetric states}

\author{Giorgio Arcadi}

\address{Laboratoire de Physique Th\'{e}orique, CNRS, Univ. Paris-Sud, Universit\'{e} 
Paris-Saclay, 91405 Orsay, France}

\ead{giorgio.arcadi@th.u-psud.fr}

\begin{abstract}
We will review the main aspects of a mechanism for the contemporary generation of the baryon and Dark Matter abundances from the out-of-equilibrium decay of a Wimp-like mother particle and briefly discuss a concrete realization in a Supersymmetric scenario. 
\end{abstract}

Preprint:LPT-Orsay-15-76

\section{Introduction}

\noindent
Two pressing puzzles of modern particle physics are the generation of the dark and visible matter components of the Universe. As further confirmed by high precision measurements by the Planck satellite~\cite{Ade:2015xua}, approximately one third of the total energy budget of the Universe is provided by a Dark Matter (DM) component, possibly represented by a cosmologically stable particle state with suppressed interactions with the visible matter. Given the absence of non-gravitational detection of DM, its particle nature and properties as well as its production mechanism are still unknown. The origin of the visible component of the Universe, contributing only to $5\%$ of the total energy budged, is similarly puzzling. Indeed, the lack of observation of primordial antimatter needs a dynamical mechanism (baryogenesis) which has originated, at some stage of the history of the Universe, an asymmetry between matter and antimatter. Although the exact mechanism is still unknown, there are three necessary conditions which should be satisfied~\cite{Sakharov:1988pm}: violation of the baryon number, violation of $CP$, departure from thermal equilibrium.  

\noindent
The solution of these two puzzles requires an extension of the Standard Model of Particle Physics (SM), since it is lacking a viable DM candidate and does not fulfill in a satisfactory way the necessary conditions for baryogenesis. A very simple possibility is to generate the baryon asymmetry and the DM abundance from the out-of-equilibrium decay of an exotic state~\cite{Arcadi:2013jza,Arcadi:2015ffa} (see also~\cite{Cui:2012jh,Sorbello:2013xwa,Cui:2013bta,Cui:2015eba} for similar ideas).

\noindent
We will provide, in this work, a pedagogical and possibly not too technical review of the general aspects of this mechanism specializing, in the last section, on the specific example of a realization of this kind of scenario in the Minimal Supersymmetric Standard Model (MSSM). 

\section{General idea}

The mechanism discussed in this work relies on the idea of generating both the DM and baryon abundances from the out-of-equilibrium decay of a particle state featuring $B$-number and $CP$ violating interactions. The two abundances can be schematically expressed as:
\begin{equation}
\label{eq:baryo_general}
\Omega_{\Delta B} =\epsilon_{\rm CP} \frac{m_p}{m_{X}}\; BR(X \rightarrow b, \bar b)\;  \Omega_{X},
\end{equation}
\begin{equation}
\label{eq:dm_general}
\Omega_{DM} =\frac{m_{DM}}{m_{X}} BR\left(X\rightarrow DM+\mbox{anything} \right) \Omega_{X}
\end{equation}
where we have used, as customary, the adimensional parameters $\Omega_i \equiv \frac{\rho_i}{\rho_c}$ with $\rho_i$ and $\rho_c$ being, respectively, the energy density of the considered species and the so called critical density.

\noindent
The two expressions depend on the initial abundance $\Omega_{\rm X}$ of the mother particle, weighted by the branching fraction in the considered channel (in the first expression $b,\bar b$ refer, respectively, to any combination of SM states with non zero baryon number and its charge conjugate; the decay into DM is assumed in general to be $B$-preserving), and the ratio between the mass of the considered species (in the case of the baryons there is the mass of the proton $m_p$) and the decaying particle. The baryon abundance depends as well on the CP-asymmetry $\epsilon_{\rm CP}$ defined as: 
\begin{equation}
\epsilon_{\rm CP} = 
\frac{\Gamma (X \rightarrow b) - \Gamma (X \rightarrow \bar b)}{\Gamma (X \rightarrow b) +\Gamma (X \rightarrow \bar b)}\; .
\end{equation}
\noindent
From the expressions above it is straightforward to evince that the correct amount of baryons and DM can be generated from a suitable initial abundance of the mother particle. In this work we will consider the case that this is a WIMP-like particle. By this we mean that the mother particle existed in thermal equilibrium at early stages of the history of the Universe after which it has undergone chemical freeze-out. The time of freeze-out and the corresponding abundance are determined by the pair annihilation rate of this particle. For conventional WIMPs the freeze-out occurs in a strongly non-relativistic regime, i.e. temperatures of the Early Universe sensitively lower than the mass of the particle, typically  $T_{\rm f.o.}=m_X/20-m_X/30$ and the relic density $\Omega_X$ is inversely proportional to the thermally averaged pair annihilation rate $\langle \sigma v \rangle$. Contrary to conventional WIMPs, the particle $X$ is, however, metastable and decays after freeze-out producing the baryon asymmetry and the DM abundance.
\noindent
The most relevant feature of this framework is that the relevant quantities are substantially determined by a particle physics inputs. As a consequence the new particles and interactions needed for the generation of the baryon and DM abundances can be potentially probed by Earth scale experiments.
\noindent
Although the expressions~(\ref{eq:baryo_general})-(\ref{eq:dm_general}) provide a very good description of the main features of the mechanism and can provide a good estimate of the considered quantities in simple models, they should be replaced by a more refined treatment once realistic particle physics frameworks are considered. This refinement consists in the numerical solution of a system of Boltzmann equations. We will briefly illustrate in the next section the main feature of this kind of system and how to quantitatively compute the DM and baryon abundances.

\section{Quantitative determination of the Baryon and DM abundances}

The simple picture depicted in the previous section is complicated by the following aspects:1) the standard approximations for the computation of $\Omega_X$ according the WIMP paradigm are problematic in the scenario under consideration. Indeed, in most particle frameworks, $\epsilon_{\rm CP}$ and $BR\left(X\rightarrow DM+\mbox{anything} \right)$ are very small numbers. They should be then compensated by a very high $\Omega_X$ which can be typically achieved considering a thermal freeze-out close to the relativistic regime. In addition, being the $X$ state not stable, interaction verteces between single $X$ states and SM particles are in general present, leading to single-annihilation processes in addition to the conventional pair annihilations, modifying the usual relation $\Omega_X \propto 1/\langle \sigma v \rangle$.2) Processes generation baryon asymmetry are typically accompanied by the so called wash-out processes. These processes also violate the baryon number and contrast the former ones tending to restore a null net baryon number. The baryogenesis mechanism should be enough efficient to compensate their effect. 3) In order to have an efficient production of the baryon asymmetry the decaying state should typically feature both sizable $B$-conserving and $B$-violating decay channels~\cite{Sorbello:2013xwa,Nanopoulos:1979gx}. The simplest way to achieve this result is introduce a third state $\psi$ lighter than the mother one, but not being the DM itself~\footnote{Possible $B$-conserving decays into DM cannot provide the desired effect because of the typically very low branching fraction.}, which triggers the $B$-conserving decays. At the same time the introduction of the new particle implies the presence of additional processes. In particular it takes a very relevant role in wash-out effects. 4) The baryon asymmetry can be as well produced in B-violating annihilations of the mother particle. 5) The decay of the mother particle might be not the only source of DM. 

\noindent
All these effects can be consistently treated by introducing three sets of coupled Boltzmann equations. The first traces the abundance of the $X$ and $\psi$ particles. The corresponding equations are schematically written as:
\begin{align}
& \frac{dY_{\tilde{\alpha}}}{dx}=-\frac{1}{H x} \Gamma_{\tilde{\alpha},\Delta B \neq 0} \left(Y_{\tilde{\alpha}}-Y^{\rm eq}_{\tilde{\alpha}}\right)-\frac{s}{H x}\langle \sigma v \rangle_{\tilde{\alpha},\Delta B \neq 0} Y_X^{\rm eq} \left(Y_{\tilde{\alpha}}-Y_{\tilde{\alpha}}^{\rm eq}\right) \nonumber\\
& -\frac{s}{Hx}\sum_{\tilde{\beta} \neq \tilde{\alpha}}\langle \sigma v \rangle \left(\tilde{\alpha}\tilde{\beta} \rightarrow X\right) \left(Y_{\tilde{\alpha}}Y_{\tilde{\beta}}-Y_{\tilde{\alpha}}^{\rm eq} Y_{\tilde{\beta}}^{\rm eq}\right)-\frac{s}{H x}\sum_{\tilde{\beta}\neq \tilde{\alpha}}\langle \sigma v \rangle \left(\tilde{\alpha}X \rightarrow \tilde{\beta} X\right)Y_X^{\rm eq}\left(Y_{\tilde{\alpha}}-\frac{Y_{\tilde{\alpha}}^{\rm eq}}{Y_{\tilde{\beta}}^{\rm eq}}Y_{\tilde{\beta}}\right)\nonumber\\
& -2 \frac{s}{H x}\langle \sigma v \rangle_{\tilde{\alpha}\tilde{\alpha}} \left(Y_{\tilde{\alpha}}^2-Y_{\tilde{\alpha}}^{\rm eq \, 2}\right) -\frac{1}{H x} \sum_{\tilde{\beta} \neq \tilde{\alpha}}\Gamma_{\Delta B = 0} \left(Y_{\tilde{\alpha}}-Y^{\rm eq}_{\tilde{\alpha}}\frac{Y_{\tilde{\beta}}}{Y_{\tilde{\beta}}^{\rm eq}}\right)\nonumber\\
& -\frac{1}{H x} \Gamma \left(\tilde{\alpha} \rightarrow DM\right) Y_{\tilde{\alpha}},\,\,\,\,\tilde{\alpha},\tilde{\beta}=X,\psi
\end{align}
where we have introduced the adimensional parameter $Y_i=n_i/s$ ($n_i$ is the number density of the considered quantity while $s$ is the entropy density) with the ratio $x=m_{\tilde{B}}/T$ as independent variable. Here the first line represents the processes originating the baryon asymmetry (it is zero for $\psi$). In general the baryon asymmetry can also be created by $B$ and $CP$ violating $2 \rightarrow 2$ scatterings. The corresponding contribution is however negligible in the specific example proposed (see instead in~\cite{Claudson:1983js,Baldes:2014gca,Baldes:2014rda} some cases in which the baryon asymmetry is mostly produced by scatterings). The second line and the first term in the third line represent the other annihilation processes, respectively with $X \psi$ as initial state (coannihilations), single $B$-conserving annihilations $X+SM \rightarrow \psi +SM$ and conventional pair annihilations. The last term is the production of Dark Matter. In concrete realizations there might be more than two particles contributing to the relevant processes. In such a case there will be a larger set of equation with the same general structure depicted above.
\noindent    
We have then the equation for the baryon asymmetry, customarily casted in terms of $Y_{\Delta B-L}$, with $L$ being the lepton number, in order to get rid of the sphaleron processes, since they conserve $B-L$.
\begin{align}
\label{eq:system_BL}
& \frac{dY_{\Delta B-L}}{dx}=\frac{1}{H x} \Delta \Gamma_{X,\Delta B \neq 0} \left(Y_X-Y_X^{\rm eq}\right)+\frac{s}{H x} \langle \Delta \sigma v \rangle_{X} \left(Y_{X}-\frac{Y_{X}^{\rm eq}}{Y_{\psi}^{\rm eq}}Y_\psi\right)\nonumber\\
& - \mbox{WO}(\mu_{\rm f})
\end{align}
Here the first line represents the sources of baryon asymmetry, with $\Delta \Gamma_{X,\Delta B}=\epsilon_{\rm CP} \Gamma_{X,\rm tot}$. The term in the second line, schematically labeled as $WO$ (the full description is given in~\cite{Arcadi:2015ffa}) represents the ensemble of wash-out processes, which, whether efficient at the time or after the generation of the baryon asymmetry, deplete the baryon abundance. The wash-out terms are proportional to the chemical potentials $\mu_f \propto n_f -n_{\bar f}$ where $f=\mbox{SM quark}$. 
The last and simplest equation is the one for the DM abundance:
\begin{equation}
\label{eq:system_gravitino}
\frac{dY_{DM}}{dx}=\frac{1}{H x} \sum_{\tilde{\alpha}} \Gamma \left(\tilde{\alpha} \rightarrow DM\right)
\end{equation}
\noindent
In general the DM is the lightest stable state of a new particle sector. As a consequence all the states belonging to this sector will decay into dark matter ($\tilde{\alpha}$ then runs over all the states of the theory). This can lead to an additional production mechanism of DM dubbed freeze-in (a complete list of references is provided in~\cite{Arcadi:2015ffa}) consisting in the production from states still in thermal equilibrium.

\noindent
This system is solved by setting the initial condition, at $x \ll 1$, $Y_{DM}=Y_{\rm \Delta B-L}=0$ and the abundances of $X$ and $\psi$ fixed to the equilibrium value, and by setting a final time $x_f$ corresponding to $x \gg x_d$ with $x_d$ being the decay time of $X$. Eq.~(\ref{eq:baryo_general})-(\ref{eq:dm_general}) are related to the solution of the Boltzmann system by $\Omega_{\Delta B, \rm DM} \propto m_{\rm p,DM} Y_{\Delta B, DM}(x_f), Y_{\Delta B}=\frac{28}{79}Y_{\Delta B-L}$.

\noindent
In the most minimal realization there are 5 relevant parameters: the three masses $m_{\rm DM},m_X,m_\psi$, the coupling $\lambda$ associated to the $B$-violating processes, and the scale $M$ which characterizes the size of the annihilation and decay processes of the $X,\psi$ particles and which can be identified as the mass of a fourth category of new particles. 

\noindent 
In the next section we will illustrate some results in a specific example. 

\section{Concrete example}

\noindent
The general framework depicted above is straightforwardly realized in the MSSM. Indeed its rich particle spectrum provide the new particles needed by the considered mechanism. Moreover baryon number is in general violated in the MSSM unless one imposes a discrete symmetry dubbed R-parity. In absence of this symmetry the superpartners feature baryon (and lepton) number violating decays into only SM states. In the example considered the Baryon number is broken by the single operator $\lambda U^c D^c D^c$, which breaks the baryon number but preserves the lepton number, avoiding conflicts with the stability of the proton. The DM candidate is the Gravitino which is not exactly stable but has a lifetime exceeding by orders of magnitude the lifetime of the Universe thanks to the Planck scale suppression of its interactions. There are, in addition, more sources of $CP$ violation with respect to the SM. The role of the $X$ state is played by the Bino while $\psi$ is one the two other gauginos. In the considered example $\psi$ is the Gluino. The role of $M$ is played by two energy scales, $m_0$ and $\mu$, representing the mass scales of the scalar superpartners and of Higgsinos. 

\noindent
The Bino-like mother particle has annihilation cross-sections going like $\langle \sigma v \rangle \propto \frac{m_{\tilde{B}}^2}{m_0^4},\frac{1}{\mu^2}$ and a total decay rate $\Gamma_{\tilde{B}}\propto \frac{m_{\tilde{B}}^5}{m_0^4}$ (the detailed expressions can be found in~\cite{Arcadi:2015ffa}). The desired value of the high abundance and long lifetime are achieved for $m_0,\mu \gtrsim 10^6\,\mbox{GeV}$ resembling the so call split and mini-split SUSY scenarios which have become rather popular on recent times since they can account the lack of detection of supersymmetric states at the LHC. 

\noindent
The Bino features R-parity violating channels into SM fermions as well as R-parity conserving decays into the Gravitino DM and the Gluino. The CP asymmetry depends on the fundamental mass scales of the theory as $\epsilon_{\rm CP} \propto \frac{m_{\tilde{B}}m_{\tilde{G}}}{m_0^2} f\left(m_{\tilde{G}}^2/m_{\tilde{B}}^2\right)$ with the function $f$ being 0 for $m_{\tilde{G}}\geq m_{\tilde{B}}$ and 1 for $m_{\tilde{G}} \ll m_{\tilde{B}}$. 

\begin{figure}[htb]
\begin{center}
\includegraphics[width=7.5 cm]{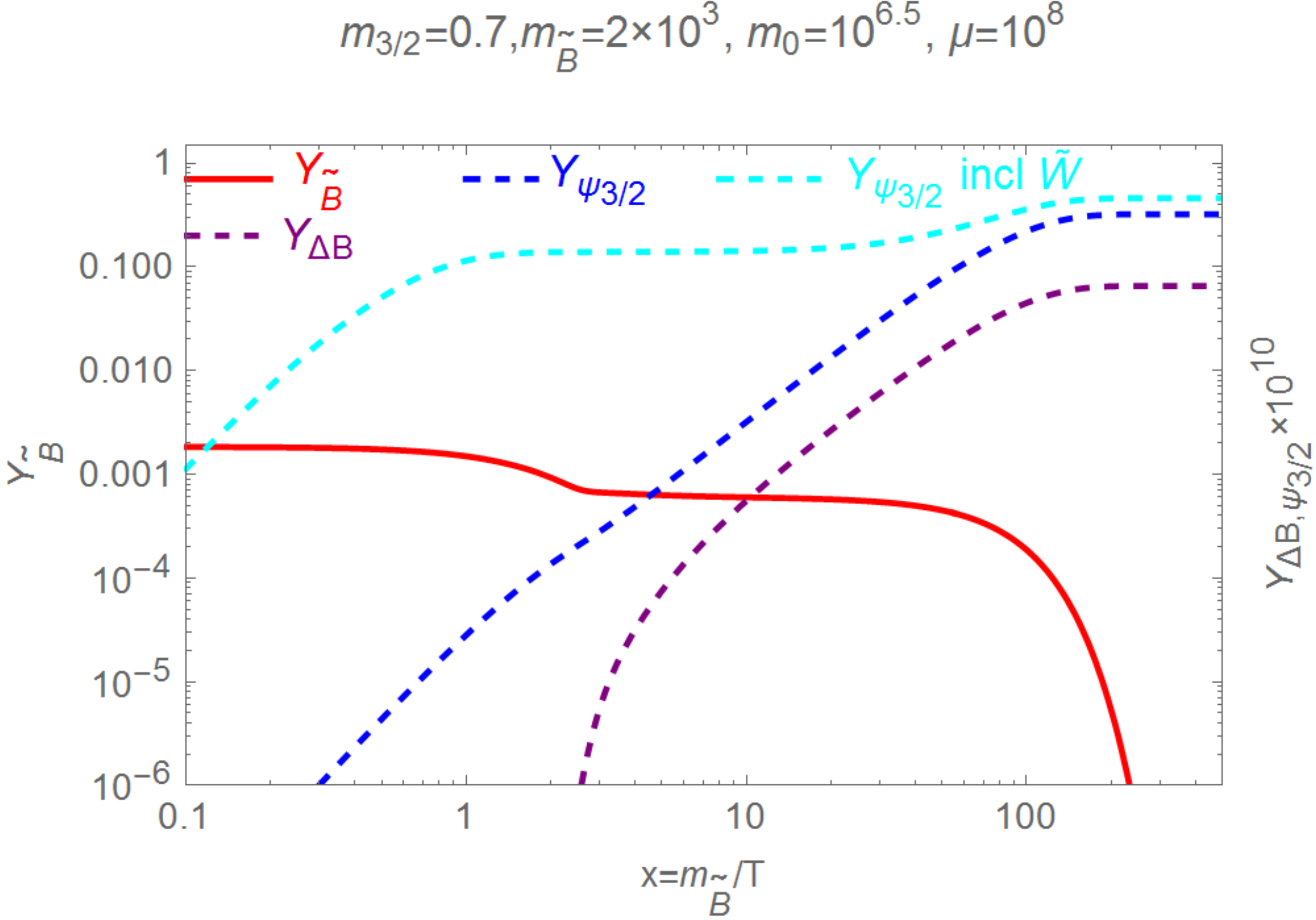}
\end{center}
\caption{\footnotesize{Sample plot showing the evolution of the abundances of the Bino (red solid curve) the baryons (purple dashed curve) and of the gravitino (blue and cyan curves). In the cyan curve is considered an additional sizable DM production from another Supersymmetric state, i.e the Wino. The assignation of the fundamental parameters is reported in the plot (the masses are in GeV).}}  
\label{fig:rainbows} 
\end{figure}

\noindent
For the details of the solution of the Boltzmann equations and the determination of the viable parameter space of this scenario we refer to the original reference~\cite{Arcadi:2015ffa}. We just report in fig.~(\ref{fig:rainbows}) a sample plot of the solution of the Boltzmann equations for a generic assignation of parameters. 

\noindent

\noindent
Here have been reported the evolution of the abundaces of the three relevant species, i.e. the Bino (mother particle), the baryon asymmetry and the gravitino DM. The purple and blue dashed lines represent the simple scenario described at the beginning in which the two abundances are contemporary generated by the Bino which occurs an almost relativistic freeze-out at $x \sim 2$ but decay at much later times. We have shown as well a possible complication due to the rich structure of Supersymmetric theory. We have indeed considered a parameter assignation in which a sizable DM component is generated by another particle of the spectrum, the Wino, heavier than the mother particle. It does not contribute at the baryogenesis process since its annihilation rate is typically very high and then its abundance after freeze-out is negligible. It can however efficiently generate the DM through the so called freeze-in mechanism (Also the Bino is capable of this kind of contribution but in this case it is neglible with respect to the out-of-equilibrium decay).

\section{Conclusions}

We have briefly reviewed the realization within the MSSM of a mechanism for the contemporary generation of the DM abundance and of the baryon asymmetry from the decay of a WIMP-like particle. This mechanism is simple and allow the prediction of the relevant quantities essentially in terms of a particle physics input. The concrete implementation in a definite particle physics framework requires a detailed numerical treatment based on the solution of suitable Boltzmann equations. We have shown an explicit implementation in a MSSM realization.

\noindent
{\bf Acknowledgements}:The author warmly thanks the organizers of the 6th Young Reseach meeting for the chance of giving this contribution and the GSSI Institute for the hospitality.
\noindent
The author is supported by the ERC advanced Grant Higgs@LHC.

\section*{References}

\bibliography{iopart-num}{}
\bibliographystyle{iopart-num}
\end{document}